\documentstyle[prd,aps]{revtex}

\input epsf

\def\notp{p\kern-4.5pt\hbox{$/$} }
\def\u#1{$\underline{\smash{\vphantom{y}\hbox{#1}}}$}
 
\title{Kaon Weak Decays in Chiral Theories\thanks{
  Published in {\it Mod.\ Phys.\ Lett.}\/ {\bf A14} (1999) 1273.}}
\author{M.D.~Scadron}
\address{Physics Dept. Univ. of Arizona, Tucson AZ  85721, USA}
\date{\today}
 
\begin{document}
\maketitle
\baselineskip 18pt

\begin{abstract}

The ten nonleptonic weak decays $K \to 2\pi$, $K \to 3\pi$, $K_L \to 2\gamma$, $K_S \to 2\gamma$, $K_L \to \pi^\circ 2\gamma$, are
predicted for a chiral pole model based on the linear sigma model theory which automatically satisfies the partial
conservation of axial current (PCAC) hypothesis.  These predictions, agreeing with data to the 
5\% level and containing no or at most one free parameter, are compared with the results of chiral perturbation theory
(ChPT).  The latter ChPT approach to one-loop level is known to contain at least four free parameters and then 
predicts a $K_L \to \pi^\circ \gamma\gamma$ rate which is 60\% shy of the experimental value.
This suggests that ChPT is an unsatisfactory approach towards
predicting kaon weak decays.

\noindent
PACs numbers:  11.30.Rd, 11.40.Ha, 13.25 Es 

\end{abstract}

\section{Introduction}

In this paper we contrast the kaon weak decay predictions of the two chiral
theories based on (i) the linear sigma model (L$\sigma$M) characterized here by the non-loop tree
graphs of the chiral pole model (CPM);  (ii)  chiral perturbation theory (ChPT) involving loop diagrams.
Prior studies of the CPM and its direct link with the model-independent approach of current algebra - partial
conservation of axial currents (PCAC) were worked out in refs.[1], while the L$\sigma$M-CPM extension
was given in ref.[2], including the weak decays $K \to 2\pi$, $K\to 3\pi$, $K_L \to \gamma\gamma$,
$K_S \to \gamma\gamma$ and $K_L \to \pi^\circ\gamma\gamma$.  At about the same time, the predictions of 
ChPT were summarized for $K \to 2 \pi$ and $K \to 3\pi$ decays in ref.[3] and extended to $K_S \to \gamma\gamma$
and $K_L \to \pi^\circ\gamma\gamma$ in ref.[4].

We shall show that the former L$\sigma$M-CPM-PCAC approach predicts the above-mentioned 10 weak decay
amplitudes to within 5\% accuracy in terms of {\em no} or at most {\em one} free parameter.  In contrast,
the latter ChPT formalism based on 10 strong interaction parameters $L_1 - L_{10}$ requires at least 4 weak
interaction parameters [3] $c_2, c_3, G_1, G_2$ to explain the 7 decays $K \to 2\pi$, $K \to 3\pi$ and
even then the one-loop ChPT prediction of the $K_L \to \pi^\circ \gamma\gamma$ rate recovers only 35\% of
the observed rate [4].

In Sec.II we study the L$\sigma$M-CPM chiral symmetry scheme for $K \to 2\pi$ and $K \to 3\pi$ decays,
predicting all 7 amplitudes in terms of tree graphs and one $\Delta I = 1/2$ scale.  The latter is at first
taken as the one fitted parameter in this scheme in Sec.II.  Then it too will be predicted from the CPM tree
approximation for $K_L \to \gamma\gamma$ in Sec.III, or from the (quark tadpole) one-loop order graph for the 
$\Delta I = 1/2$ $s \to d$ self energy in Sec.IV.  Also in Sec.III we extend this L$\sigma$M-CPM chiral symmetry
approach to tree graphs for $K_S \to \gamma\gamma$ and $K_L \to \pi^\circ\gamma\gamma$.  Finally in Sec.V we summarize
the ChPT results for the 10 weak decays and indicate where {\em two} $\Delta I = 1/2$  and {\em two}  independent $\Delta I = 3/2$ 
fitted parameters and also {\em one} $K_L \to \pi^\circ\gamma\gamma$ fitted parameter are required.  We draw our
conclusions in Sec.VI.

\section{L$\sigma$M-CPM-PCAC approach to $K \to 2\pi$ and $K \to 3\pi$ decays}

The strong interaction SU(2) linear $\sigma$ model (L$\sigma$M) lagrangian and its implication for chiral symmetry and
partial conservation of axial currents (PCAC) are well-documented in text books [5].  The natural extension of the SU(2)
L$\sigma$M (for pseudoscalar $\vec{\pi}$ and scalar $\sigma$ mesons) to weak interactions of kaons is via a
chiral pole model (CPM) involving again intermediate $\vec{\pi}$ and  $\sigma$ mesons [2,6].

Specifically the dominant $\Delta I = 1/2$ CPM graph is depicted in Fig. 1 for parity-violating (pv) 
$K_S \to \pi\pi$ decays via $K_S^{PV} \to \sigma \to 2\pi$, with the latter $\sigma \to \pi\pi$ transition given by the 
L$\sigma$M vertex [5] $\langle \pi\pi | \sigma \rangle = - m^2_\sigma / f_\pi$ for $f_\pi \approx$ 93 MeV.  The former
weak vertex $\langle \sigma | H_w^{pv} | K_S \rangle $ is given by the chiral symmetry relation
$$
\langle \sigma | H_w^{pv} | K_S \rangle  = \langle \pi^\circ | H_W^{pc} | K_L \rangle .
\eqno(1)
$$
Since the intermediate $\sigma$ resonance has a broad width as suggested by many experiments [7], or from the
L$\sigma$M theory or mended chiral symmetry [8] with $\Gamma_\sigma \approx m_\sigma \sim 700$ MeV, the
$\Delta I = 1/2$ CPM $K_S \to 2\pi$ amplitude in the chiral limit based on Fig. 1 is [1,2]
$$
\langle \pi \pi | H^{pv}_w|K_S\rangle = \langle\pi\pi | \sigma\rangle {1 \over
m^2_K - m^2_\sigma + i m_\sigma \Gamma_\sigma} \langle \sigma | H^{pv}_w|K_L\rangle $$ $$ 
\approx (i/f_\pi ) \langle \pi^\circ | H^{pc}_w | K_L \rangle 
\eqno(2)
$$
when $m_\pi = 0$.
Here we have used (1) and dropped the small real part of (2)  relative to its imaginary part since 
$|m^2_K - m^2_\sigma| << m^2_\sigma$.  This L$\sigma$M-CPM result (2) also is a consequence [1] of PCAC applied
to both pions (PCAC consistency) with charge communtator amplitude $M_{CC}:$
$$
\langle \pi_1\pi_2 | H_w | K_S \rangle  = M_{CC1} + M_{CC2} + {\cal O}(m^2_\pi /m^2_K ). 
\eqno(3)
$$

Returning to the CPM version (2), the value $\langle \pi^\circ | H^{pc}_w | K_L \rangle | \approx 3.2 \times
10^{-8}\ GeV^2$ to be found in Sec.III from the CPM version of $K_L \to \gamma\gamma$ in turn sets the 
$K_S \to 2\pi^\circ$ $\Delta I = 1/2$ scale from eq.(2) for $f_\pi \approx 93$ MeV: 
$$
|\langle \pi^\circ \pi^\circ | H^{pv}_w|K_S\rangle|_{CPM} \approx |\langle \pi^\circ  | H^{pc}_w|K_L\rangle| / f_\pi
\approx 34 \times 10^{-8} \ GeV.
\eqno(4)
$$
The CPM extension to $K_S \to \pi^+\pi^-$ includes Fig. 1 along with Fig. 2 for charged pions.  These
latter W emission graphs ($W_{em})$ have small
$\Delta I = 1/2$  and $\Delta I = 3/2$ parts and can be computed using the ``vacuum
saturation" method [9]
$$
|\langle \pi^+ \pi^- | H^{pv}_w|K_S\rangle|_{Wem} 
= (G_F s_1c_1/ 2\sqrt{2} ) | \langle \pi^+ | A_\mu | 0 \rangle 
\langle \pi^- | V^\mu | K_S \rangle | + \leftrightarrow 
$$ $$
\hskip 23mm =G_F s_1c_1 f_+ (0) f_\pi (m^2_K - m^2_\pi)/\sqrt{2}  \approx 4 \times 10^{-8} \ GeV,
\eqno(5)
$$
for V-A chiral left-handed vector currents simulating the vector W. Then  the total $K_S \rightarrow
\pi^+\pi^-$ weak CPM amplitude is the sum of (4) and (5):
$$
|\langle \pi^+ \pi^- | H^{pv}_w|K_S\rangle|_{CPM}
\approx (34 + 4) \times 10^{-8} \ GeV = 38 \times 10^{-8} \ GeV.
\eqno(6)
$$
Lastly the pure $\Delta I = 3/2$  $K^+ \to \pi^+\pi^\circ$ amplitude can be computed in the CPM via the analog W
emission (or vacuum saturation) value [9]
$$
|\langle \pi^+ \pi^\circ | H^{pv}_w|K^+\rangle|_{CPM}
= (G_F s_1c_1/ 2\sqrt{2} ) | \langle \pi^+ | A_\mu | 0 \rangle 
\langle \pi^\circ | V^\mu | K^+ \rangle |
$$ $$
\hskip 23mm =G_F s_1c_1 f_+(0) f_\pi(m^2_K - m^2_\pi)/2\sqrt{2}  \approx 1.83 \times 10^{-8} \ GeV.
\eqno(7)
$$
In (5) and (7) we invoke $f_+(0) \approx 0.96 $ as the ${\cal O} (\varepsilon^2)$ small deviation from the
nonrenormalization limit of unity as found in various quark model schemes [10].

Although the above CPM is quite simple (yet manifesting chiral symmetry), it is also very accurate as the 
following experimental (exp) amplitudes $M^{\pi\pi}$ indicate [11]:
$$
|M^{+-}_{K_S} |_{exp} = (39.08 \pm 0.08 ) \times 10^{-8} \ GeV
$$
$$
|M^{00}_{K_S} |_{exp} = (37.11 \pm 0.17 ) \times 10^{-8} \ GeV
$$
$$
|M^{+0}_{K^+} |_{exp} = (1.833 \pm 0.006 ) \times 10^{-8} \ GeV.
\eqno(8)
$$
The CPM predictions (4), (6), (7) are respectively within 1\%, 8\%, 1\% of the observed $K_{2\pi}$ amplitudes in (8).

Similar 5\% accuracy for these $K_{2\pi}$ $\Delta I = 1/2$ and $\Delta I = 3/2$ scales follows by invoking ``PCAC 
consistency" [1] of eq. (3), giving 
$$
a^{+-}_S = i \langle \pi^+ \pi^- | H_w|K_S\rangle = \langle \pi^+  | H_w|K^+\rangle
(1-m^2_\pi / m^2_K) / f_\pi
$$
$$
a^{00}_S = i \langle \pi^0 \pi^0 | H_w|K_S\rangle = \langle \pi^0  | H_w|K_L\rangle
(1-m^2_\pi / m^2_K) / f_\pi
$$
$$
a^{+0}_+ = i \langle \pi^+ \pi^- | H_w|K^+\rangle = \langle \pi^+  | H_w|K^+\rangle
(1-m^2_\pi / m^2_K) / 2f_\pi
$$
$$
+ \sqrt{2} \langle \pi^\circ  | H_w|K^\circ\rangle (1-m^2_\pi / m^2_K) / 2f_\pi.
\eqno(9)
$$
Note that the $a^{00}_S$ equation is compatible with CPM-PCAC given by (2).  Note too the explicit factors of 
$(1-m^2_\pi / m^2_K)$ occurring in eqs.(9) which force all $K_{2\pi}$ amplitudes to vanish in the strict SU(3)
limit, a result originally obtained by Cabibbo and Gell-Mann [12] due to CP and SU(3) invariance.
Then one models the reduced matrix elements $\langle\pi^+  | H_w|K^+\rangle$ and $\langle \pi^0  | H_w|K^\circ\rangle$
via the s-d quark self energy and the W-exchange graphs [2,13] or alternatively uses a pure meson loop model
[1].  Lastly one can further tune the above 5\% CPM discrepancy to the 2\% level by accounting for final-state
$\pi\pi$ interactions [1,13] given the observed $\delta_0 - \delta_2 \approx 57^\circ$ phase shift difference, but
we shall not do so here.

Instead we accept the above CPM predictions for the three $K_{2\pi}$ amplitudes to 5\% accuracy (but containing {\em no}
free parameters), and extend the scheme to the four $K_{3\pi}$ amplitudes via PCAC consistency [1,13] in analogy with (2) and (9):
$$
A^{+-0}_L = i \langle \pi^+ \pi^-\pi^0 | H_w|K_L\rangle = - \langle \pi^0 \pi^0 | H_w|K_S\rangle (1-m^2_\pi / m^2_K) / 4f_\pi 
$$
$$
A^{00+}_L = i \langle \pi^0 \pi^0 \pi^+| H_w|K^+\rangle =  \langle \pi^+ \pi^- | H_w|K_S\rangle (1-m^2_\pi / m^2_K) / 4f_\pi 
$$
$$
A^{++-}_L = i \langle \pi^+ \pi^+ \pi^-| H_w|K^+\rangle = 2 \langle \pi^+ \pi^- | H_w|K_S\rangle (1-m^2_\pi / m^2_K) / 4f_\pi 
$$
$$
A^{000}_L = i \langle \pi^0 \pi^0 \pi^0| H_w|K_L\rangle = -3 \langle \pi^0 \pi^0 | H_w|K_S\rangle (1-m^2_\pi / m^2_K) / 4f_\pi .
\eqno(10)
$$
In the final forms of eqs.(9) we have used the $K_{2\pi}$ sum rule $M^{+-}_S - M^{00}_S = 2 M^{+0}_+$ along with the
PCAC consistency extension of $K_{2\pi}$ in (3) to the $K_{3\pi}$ version [1]
$$
\langle \pi_1 \pi_2 \pi_3 | H_w | K \rangle = {\tiny {1 \over 2}} (M_{CC1} + M_{CC2} + M_{CC3} ) + {\cal O} (m^2_\pi/m^2_K).
\eqno(11)
$$
The factor of ${1 \over 2}$ in (11) (already occurring in (10)) accounts for the ``mismatch" between Feynman amplitudes
(where the pions are treated as independent) and PCAC consistency (where the PCAC procedure must be symmetrized over the
final-state pions) with the decaying kaon always kept on mass shell.
Just as the PCAC consistency $K_{2\pi}$ form (3) also follows from a (tedious) analysis of rapidly varying pole terms
[1], the PCAC consistency $K_{3\pi}$ form (11) (including the factor of 1/2) likewise follows from an (even more tedious) analysis
of rapidly varying pole terms [14].

Given the three $K_{2\pi}$ L$\sigma$M-CPM predictions, (4), (6), (7), the PCAC consistency extension to the four
$K_{3\pi}$ amplitudes in (10) is 
$$
|A^{+-0}_L |_{PCAC} \approx 0.85 \times 10^{-6}
$$
$$
|A^{00+}_+ |_{PCAC} \approx 0.95 \times 10^{-6}
$$
$$
|A^{++-}_+ |_{PCAC} \approx 1.89 \times 10^{-6}
$$
$$
|A^{000}_L |_{PCAC} \approx 2.53 \times 10^{-6}.
\eqno(12)
$$
These $K_{3\pi}$ predictions in (12) are respectively within 6\%, 1\%, 2\%, 3\% of the experimental amplitudes
[11]
$$
|A^{+-0}_L |_{exp} = (0.91 \pm 0.01) \times 10^{-6}
$$
$$
|A^{00+}_+ |_{exp} = (0.96 \pm 0.01) \times 10^{-6}
$$
$$
|A^{++-}_+ |_{exp} = (1.93 \pm 0.01) \times 10^{-6}
$$
$$
|A^{000}_L |_{exp} = (2.60 \pm 0.02) \times 10^{-6}.
\eqno(13)
$$
The latter  amplitudes are extracted from the standard three-body phase space integral [15] with $N$ being the
Feynman statistical factor for the rate
$$
\Gamma = [ 2/N (8\pi M)^3 ] |A|^2 \int^{(M-m)^2}_{4\mu^2} ds \left[ { [s-4\mu^2 ] [s - (M+m)^2]
[s-(M-m)^2] \over s} \right]^{1/2}
$$
$$
= I |A|^2 ,
\eqno(14)
$$
where M is the kaon mass, $m$ is the odd-pion mass, and $\mu$ is the non-odd-pion mass.  The amplitudes
$A$ in (14) are taken as constant (empirically valid to within 5\% ) and the resulting integrals in (14) are
$I(+-0) = 1.95,$ $I(00+) = 0.996,$ $I(++-) = 0.798,$ $I(000) = 0.397$ in units of $10^{-6}\ GeV$.

Suffice it to say that this L$\sigma$M-CPM-PCAC approach used in Sec. II predicts all {\em seven} $K_{2\pi}$
and $K_{3\pi}$ weak decay amplitudes to within 5\% accuracy relative to the data - in terms of just {\em one}
$\Delta I = 1/2$ scale here
$$
| \langle \sigma | H_w | K_S \rangle | = | \langle \pi^\circ | H_w | K_L \rangle | \approx
3.2 \times 10^{-8} \ GeV^2.
\eqno(15)
$$

\bigskip
\bigskip
\bigskip

\section{Extension of CPM to $K_L \to \gamma\gamma, K_S \to \gamma\gamma, K_L \to \pi^\circ \gamma\gamma$}

First we consider $K_L \to 2\gamma$ decay with CPM $\pi^\circ$ pole graph of Fig.3 generating the amplitude
$$
\langle 2 \gamma | H^{pc}_w | K_L \rangle = \langle 2 \gamma |\pi^\circ\rangle (m^2_K - m^2_\pi)^{-1}
\langle \pi^\circ | H^{pc}_w | K_L \rangle
= F_{K_L \gamma\gamma} \varepsilon^{\prime\mu} \varepsilon^\nu \varepsilon_{\mu\nu\alpha\beta}
k^{\prime\alpha} k^\beta.
\eqno(16)
$$
One knows the ABJ [16] or equivalently the L$\sigma$M $\pi^\circ \to \gamma\gamma$ amplitude has magnitude
$\alpha / \pi f_\pi$ and the analogue $F_{K_L \gamma\gamma}$ amplitude for branching ratio [11] $5.9 \times
10^{-4}$ with lifetime $\tau_{K_L} = 5.17 \times 10^{-8}$ sec. gives
$$
|F_{K_L \gamma\gamma}| = \left[ {64 \pi \over m^3_K} \Gamma_{K_L\gamma\gamma} \right]^{1/2} = 
(3.51 \pm 0.04) \times 10^{-9} GeV^{-1}. 
\eqno(17)
$$
Then eq.(16) requires the scale
$$
|\langle \pi^\circ | H_w | K_L \rangle| \approx 3.2 \times 10^{-8} \ GeV^2 ,
\eqno(18)
$$
which matches the $\Delta I = 1/2$ scale of (15) needed to explain all $K_{2\pi}$ and $K_{3\pi}$
decays by construction.  

Next we apply the CPM and the $\sigma$ pole graph of Fig 4 to compute the $K_S \to \gamma\gamma$ decay
amplitude [2]
$$
\langle 2 \gamma | H^{pv}_w | K_L \rangle = F_{K_S \gamma\gamma} \varepsilon^{\prime\mu} \varepsilon^\nu 
(k_\mu k^\prime_\nu - k^\prime k g_{\mu\nu} )
$$
$$
\hskip 15mm = \langle 2\gamma | \sigma \rangle (m^2_K - m^2_\sigma + i m_\sigma \Gamma_\sigma )^{-1}
\langle  \sigma | H^{pv}_w | K_S \rangle.
\eqno(19)
$$
The scalar analogue $\sigma \to 2 \gamma$ of the L$\sigma$M $\pi^\circ \gamma\gamma$ amplitude
in (17) receives a quark-loop $u$ and $d$ enhancement of 5/3 in Fig 5a:
$$
F_{qk\ loop} = N_c ({4 \over 9} + {1 \over 9} ) {\alpha \over \pi f_\pi} = {5 \over 3} {\alpha
\over \pi f_\pi}
\eqno(20)
$$
for $N_c = 3$.  But the L$\sigma$M also requires the $\pi^+$ meson loop of Fig 5b, generating the $\sigma 
\to \gamma\gamma$ amplitude [17]
$$
F_{\pi\ loop} = - {2g^\prime \alpha \over \pi m^2_\sigma } \left[ - {1 \over 2} + \xi I (\xi ) \right] =
- \left[ - {1 \over 2} + \xi I (\xi ) \right] {\alpha\over \pi f_\pi},
\eqno(21)
$$
where we have used the L$\sigma$M coupling $g^\prime = m^2_\sigma / 2f_\pi$.  With $\xi \equiv m^2_\pi /
m^2_\sigma \approx 0.04 $ for [7,8] $\sigma (700)$, the Feynman integral $I(\xi )$ in (21) is [17]
$$
I(\xi ) = \int^1_0 dy \ y \int^1_0 dx \ \left[ \xi - xy(1-y) \right]^{-1} =
{\pi^2 \over 2} - 2 \ln^2 \left[ { 1 \over \sqrt{4 \xi} } + \sqrt{{ 1 \over 4 \xi } - 1} \right]
\approx .025.
\eqno(22)
$$
Substituting (22) into (21), one notes that the pion loop amplitude of Fig 5b changes sign [18] and enhances
the quark loop amplitude of (20), giving for (21)
$$
F_{\pi\ loop} = - (-0.50) {\alpha \over \pi f_\pi} .
\eqno(23)
$$
Then the net SU(2) L$\sigma$M $\sigma \to \gamma\gamma$ amplitude is 
$$
F^{L\sigma M}_{\sigma\gamma\gamma} \approx (1.67 + 0.50 ) {\alpha \over \pi f_\pi} = 2.17 {\alpha \over \pi f_\pi},
\eqno(24)
$$
predicting a scalar $\to \gamma\gamma$ rate now compatible
with data [19].

Returning to the $K_S \to \gamma\gamma$ amplitude  (19) and using the same approximation $|m^2_K - m^2_\sigma | << 
m^2_\sigma$ as in (2) we find, given the observed [11] branching ratio $B(K_S \to \gamma\gamma) = (2.4 \pm 0.9 )
\times 10^{-6}$ and corresponding amplitude $F_{K_S\gamma\gamma}=(5.4 \pm 1.0) \times 10^{-9}\  GeV$, 
$$
|\langle \sigma | H_w | K_S \rangle| \approx m^2_\sigma | { F_{K_S\gamma\gamma} \over F^{L\sigma M}_{\sigma\gamma\gamma}}
| = (4.9 \pm 0.9) \times 10^{-8} \ GeV^2
\eqno(25)
$$
assuming $m_\sigma \approx 700 \ MeV$.  Actually we prefer [20] the L$\sigma$M-NJL scalar mass
$m_\sigma = 2 m_q \approx 650 \  MeV$, in which case (25) predicts $|\langle \sigma | H_w | K_S \rangle| 
= (4.2 \pm 0.8) \times 10^{-8} \ GeV^2$

Although the latter estimate is within one standard deviation of the $K_L \to \gamma\gamma$ value (18) for this 
crucial $\Delta I = 1/2$ weak scale, the extreme sensitivity of (25) on $m^2_\sigma$ makes this latter 
successful estimate at best only plausible (but nonetheless consistent with the overall CPM picture).  Stated in a 
more  phenomenological way, a CPM picture for $K_S \to \gamma\gamma$ decay dominated by an  intermediate scalar
$\varepsilon (1000)$ with observed PDG rate [11] $\Gamma_{\varepsilon\gamma\gamma} \approx  6 \ keV$ (as emphasized
in ref.[19]) roughly predicts a $K_S \to \gamma\gamma$ rate
$$
\Gamma_{K_S\gamma\gamma} \sim \Gamma_{\varepsilon\gamma\gamma}  
|\langle \varepsilon| H_w | K_S \rangle|^2 /m^4_\varepsilon \sim 6 \times 10^{-21} \ GeV
\eqno(26)
$$
for our usual $\Delta I = 1/2$ weak scale $|\langle \varepsilon| H_w | K_S \rangle| \sim 3.2 \times 10^{-8} \ GeV^2$
as given by (15) or (18).  For this rate (26) to be compatible with data [11],
$$
\Gamma_{K_S\gamma\gamma} = (18 \pm 7) \times 10^{-21} \ GeV,
\eqno(27)
$$
the scalar mass $\varepsilon$(1000) in (26) should be replaced by $\sigma$(760), close to the theoretical
value in ref.[20].

Finally we study $K_L \to \pi^\circ \gamma\gamma$ in the CPM.  Following ref.[2] we consider only the
CPM graph of Fig 6, generating the weak parity-conserving (pc) amplitude
$$
\langle \pi^\circ\gamma_{q_1} \gamma_{q_2} | H^{pc}_w | K_L \rangle = \langle \gamma\gamma|\sigma\rangle
{ 1 \over s-m^2_\sigma+i m_\sigma \Gamma_\sigma } \langle \pi^\circ\sigma | H^{pc}_w | K_L \rangle
\eqno(28)
$$
where $s = (q_1 + q_2)^2$.  We shall use the chiral symmetry constraint analogous to eq(1):
$$
\langle \pi^\circ\sigma | H^{pc}_w | K_L \rangle = \langle \pi^\circ\pi^\circ| H^{pc}_w | K_S \rangle
\eqno(29)
$$
and scale the latter directly to $K_S \to \pi^\circ \pi^\circ$ data in eq.(8) (or equivalently the predicted CPM 
amplitude in eq.(4)).  The corresponding weak decay rate involves the three-body phase space integral
[15,21] over the square of (28):
$$
\Gamma ( K_L \to \pi^\circ \gamma\gamma) = |\langle \pi^\circ\sigma | H_w | K_L \rangle |^2
|F_{\sigma\gamma\gamma}|^2 {\pi^2 \over m^3_K (4\pi)^5} \times
$$
$$
\int^{(m_K - m_\pi)^2}_{s_\circ} \ \ ds\ s^2 { \{ [ s-(m_K + m_\pi)^2][s-(m_K - m_\pi)^2]\}^{1/2} \over 
(s - m^2_\sigma)^2 + m^4_\sigma }.
\eqno(30)
$$
The integral in (30) has the numerical value $1.7 \times 10^{-4} \ GeV^4$ for the same lower cutoff
$s_\circ = 0.0784 \ GeV^2$ as used by the experimental groups [22] which measured the rate of 
$K_L \to \pi^\circ \gamma\gamma$, the latter PDG average being [11]
$$
\Gamma ( K_L \to \pi^\circ \gamma\gamma)_{exp} = (2.16 \pm 0.36) \times 10^{-23}\ GeV.
\eqno(31)
$$
Using the chiral symmetry relation (29), the CPM prediction (4) (only 5\% shy of the observed $K_{2\pi^\circ}$
amplitude), and the L$\sigma$M $\sigma \to 2 \gamma$ amplitude (24) (only 10\% shy of the data [19]), the
predicted CPM rate in (30) becomes
$$
\Gamma ( K_L \to \pi^\circ \gamma\gamma)_{CPM} =  
|\langle 2\pi^\circ | H_w | K_S \rangle |^2
(2.17 {\alpha \over \pi f_\pi})^2 ({\pi^2 \over m^3_K (4\pi)^5})
\times $$ $$ \hskip 15mm (1.7 \times 10^{-4} \ GeV^4) \approx 1.5 \times 10^{-23}\ GeV,
\eqno(32)
$$
within 2 standard deviations of the measured rate in (31).  Moreover, the CPM invariant $\gamma\gamma$
spectrum in Fig. 6 of ref.[2] peaks in a manner compatible with data, a result also true for ChPT [4,21].

Thus the 3 weak radiative rates computed in this section III for $K_L \to \gamma\gamma$, $K_S \to 2\gamma$, 
$K_L \to \pi^\circ 2 \gamma$ have the CPM predictions in (18), (25 or 26), (32) which are all near the data in (17),
(27), (31), respectively.

\section{Single Quark Line Prediction for $\Delta I = 1/2$ Scale}

To complete the L$\sigma$M-CPM picture, we should reconfirm this one $\Delta I = 1/2$ scale based on the underlying
quark model, where e.g. the quark loop for $\pi^\circ \to 2 \gamma$ or its extension to $\sigma \to 2 \gamma$ do make contact with
data.  To this end we consider the $\Delta I = 1/2$ single quark line (SQL) transition $s \to d$ depicted in Fig.7 via the self energy
effective hamiltonian $\Sigma_{sd} = b \bar{d} \notp(1 - \gamma_5)s + h.c.$ according to the dimensionless weak scale [23]
$$
-b \approx {G_F s_1 c_1 \over 8 \pi^2 \sqrt{2} } (m^2_c - m^2_u) \approx 5.6 \times 10^{-8}.
\eqno(33)
$$
Here the GIM [24] enhancement factor $m^2_c - m^2_u$ in (33) is big because the charmed quark mass $m_c \approx
1.6\ GeV$ is large relative to $m_u \approx 0.34\ GeV$.

Recently it has been shown [25] that this SQL $\Delta I = 1/2$ scale (33) not only predicts $K_S \to \pi\pi$
correctly, but it also maps out hyperon $B \to B^\prime \pi$, $\Xi^- \to \Sigma^- \gamma$ and $\Omega^- \to \Xi^- 
\gamma$ weak decays.  It is sometimes suggested that this SQL scale (33) can be transformed away for
$K_S \to \pi\pi$ decays.  While we have previously argued that this  
cannot be done for $K_{2\pi}$ decays [26], it most certainly cannot be extended to the above SQL hyperon decays in
any case (else these hyperon decays would vanish).  Thus we proceed with (33) and apply it to $K_{2\pi}$ decays.

Specifically the first-order weak axial-vector LSZ amplitude is [27]
$$
M_\mu = i \int d^4x \ e^{iqx} \langle 0 | T(H^{pc}_w A^3_\mu (x)) | K^\circ \rangle \approx 
ib\sqrt{2} f_K q_\mu,
\eqno(34)
$$
where the weak scale $b$ multiplies the strong axial current as depicted in Fig.8.  This multiplication suggests
a very short-distance weak structure of (33) relative to the strong scale generating $f_K$ (because $M^{-1}_W <<
m^{-1}_K$).  Then the soft-pion theorem predicts on the kaon mass shell
$$
q^\mu M_\mu = i f_\pi \langle \pi^\circ | H^{pc}_w  | K^\circ \rangle = ib\sqrt{2} f_K m^2_K
\eqno(35)
$$
$$
\langle \pi^\circ | H^{pc}_w  | K_L \rangle = 2b (f_K / f_\pi) m^2_K \approx -3.4 \times 10^{-8}\ \ GeV^2.
\eqno(36)
$$
for $(f_K / f_\pi)\approx 1.2$ and $ b \approx -5.6 \times 10^{-8}$ from (33).

We note that this predicted $\Delta I = 1/2$ SQL scale in (36) is very close to the $3.2 \times 10^{-8}\ \ GeV^2$ scale
in (15) and (18) needed to properly fix the $K_L \to 2 \gamma$ rate.  If instead we fixed the 
$\langle \pi^\circ | H_w  | K_L\rangle$ scale in (15) and (18) to this predicted SQL-GIM-enhanced scale of (36) driven
by (33), then the ``worst" $K_{2\pi}$ and $K_{3\pi}$ CPM predictions in (4) for $K^\circ_{2\pi}$ and in (12) for $A^{+-0}$
become even closer to the data, namely 1\% and 2\% respectively.

\section{Chiral Perturbation Theory Predictions}

In ref.[3] it was shown that the three $ K_{2\pi}$ amplitudes could be accurately predicted if two parameters, $c_2$ for
$\Delta I = 1/2$ and $c_3$ for $\Delta I = 3/2$ transitions, were allowed to be fitted freely.  Moreover, higher
order four-derivative couplings (generating 82 terms) are needed in ChPT to explain the four $K_{3\pi}$ amplitudes to
within 5\%.  This corresponds to fitting not only $c_2$ and $c_3$ (as in $K_{2\pi}$ decays), but also two more
parameters $G_1$ and $G_2$.

Then in ref.[4] the $K_S \to 2 \gamma$ and $K_L \to \pi^\circ 2 \gamma$ decays were considered (but not
$K_L \to 2 \gamma$).  For $K_S \to 2 \gamma$ the tree-level and one-loop level ChPT theory predictions
(generating 37 terms in the four-derivative Lagrangian)  are in good agreement with the branching ratio 
$B(K_S \to \gamma\gamma) = 2.0 \times 10^{-6}$ (near the PDG value $(2.4 \pm 0.9) \times 10^{-6})$ 
{\em provided}  the parameter $G^{CA}_8$ is freely fitted to  $9.1 \times 10^{-6} \  GeV^{-2}$.  Given this value of $G^{CA}_8$,
the resulting $K_L \rightarrow \pi^\circ 2 \gamma$ rate in one-loop order ChPT has branching ratio $0.68 \times 10^{-6}$,
which is only $40\%$ of the observed  $K_L \rightarrow \pi^\circ 2 \gamma$ branching ratio [15] of $1.70 \times 10^{-6}$.
However as noted before, the ChPT $\gamma\gamma$ spectrum for $K_L \rightarrow \pi^\circ 2 \gamma$ roughly matches
the data, as does the L$\sigma$M-CPM $\gamma\gamma$ spectrum.

In Table 1 we contrast the predictions of the L$\sigma$M-CPM-PCAC approach described in Secs II-IV with the one loop
ChPT results summarized in Sec.V and compare them to experiment.

\centerline{\bf Table 1:  Contrasting Chiral Theories}

\hskip 20mm \u{L$\sigma$M-CPM-PCAC} \hskip 32mm \u{ChPT}

$$
\begin{array}{lll}
K \to 2\pi\ \ \ \ \ & {\rm Predicts\ all\ 3\ amplitudes} & {\rm Two\ fitted} \\
\ &{\rm to\ within \ 5\%\ of \ data\ with}\ \ \ \ \ \ \ \ \ \ \ & {\rm parameters}\ c_2,\  c_3 \\
\ &          {\rm no \ free \ parameters} & \  \\
K \to 3\pi & {\rm Predicts\ all\ 4\ amplitudes} & {\rm Four\ fitted} \\
\ &          {\rm to\ within \ 5\%\ of \ data\ with} & {\rm parameters}\ c_2,\  c_3 \\ 
\ &          {\rm no \ free \ parameters} & G_1, \ G_2  \\
K_L \to 2 \gamma & {\rm Amplitude\ predicted\ to \ within} & ?  \\
\ & 3\% {\rm\ of\ data}& \  \\
K_S \to 2 \gamma & {\rm Amplitude\ predicted\ to \ within} & {\rm One\ fitted\ parameter}  \\
\ & 15\%  {\rm\ of\ data} & G^{CA}_8  \\
K_L \to \pi^\circ 2\gamma & {\rm Rate\ predicted\ to \ within} & {\rm Given\ }G^{CA}_8\ {\rm above,\ predicts}  \\
 \ & {\rm 28\% \  of\ data}& {\rm branching\ ratio\ 40\%\ of\ data} \\
\end{array}
$$

\newpage
\section{Conclusion}

In this paper we have shown that the chiral symmetry approach of the SU(2) linear $\sigma$ model (L$\sigma$M) 
extended for weak interactions to the chiral pole model (CPM), involving tree-level $\pi^\circ$ and 
$\sigma$ poles, provides a very accurate description of nonleptonic weak kaon decays.  Specifically if we input the
one $\Delta I  = 1/2$ scale derived from a single quark line (SQL) GIM-enhanced transition nonperturbatively
inducing
$$
- \langle \pi^\circ | H_w | K_L \rangle = {G_F s_1 c_1 \over 4 \pi^2 \sqrt{2} } (m^2_c - m^2_u) (f_K / f_\pi)
m^2_K \approx 3.4 \times 10^{-8} \ GeV^2,
\eqno(37)
$$
then the 8 predicted decays $K \rightarrow 2\pi$, $K \to 3\pi$, $K_L \to 2 \gamma$ all match experiment to within
2\% - without introducing any free parameters.  Moreover the decays $K_S \to 2 \gamma$, $K_L \to \pi^\circ 2 \gamma$
are then predicted to be within 2 standard deviations of the data central values scaled
to this weak SQL transition (37).  At the very least, even if the SQL scale (37) is not used, then this L$\sigma$M-CPM-PCAC
scheme correctly predicts these 10 decay amplitudes in terms of only {\em one} free parameter.

Since this $K_{2\pi}$ L$\sigma$M-CPM scheme reduces to standard PCAC formulae, we have also used PCAC to obtain
our $K_{3\pi}$ predictions.  By way of contrast we have compared the above L$\sigma$M-CPM-PCAC results with the
much more complicated and far less predictive approach of chiral perturbation theory (ChPT).

In particular, the two scales of $K_{2\pi}$ decays, for $\Delta I = 1/2$ and for $\Delta I = 3/2$ transitions, must
both be assumed for ChPT (whereas they are both predicted accurately in the L$\sigma$M-CPM-PCAC scheme).  Furthermore 
two more ChPT parameters must be assumed for $K_{3\pi}$ decays (even with the cumbersome 82 Lagrangian terms).  Moreover
the single $K_S \to 2 \gamma$ weak scale must be assumed (even with 37 more terms in the Lagrangian), and then the 
$K_L \to \pi^\circ 2 \gamma$ ChPT rate is only 40\% of the data.

We therefore conclude that the former L$\sigma$M-CPM-PCAC chiral symmetry approach is far more predictive and less
complicated than is ChPT.  In a prior study [28] we also conclude that a L$\sigma$M approach to pion interactions occurring
in strong transitions, $r_\pi$, $F_A(0)/F_V(0)$, $\alpha_{\pi +}$, $a^{(0)}_{\pi\pi}$ is also more predictive than is
ChPT.  

It is interesting that there has been a recent attempt [29] to merge a L$\sigma$M-type picture with 
$m_\sigma \sim 700$ MeV together with $K \to 2 \pi$ weak decays and ChPT.  While this former link is compatible
with data and with refs.[1] and [2], the above analysis suggests that an extension to ChPT is quite implausible.

The author appreciates discussions with A. Bramon, S. Choudhury, R. Delbourgo, V. Elias, and
R. Karlsen.  

\section{Figure Captions}

\begin{description}

\item{Fig.1}  CPM graph for $\Delta I = 1/2$ $K_S \to \pi\pi$ amplitudes.

\item{Fig.2}  W-emission extension to $\Delta I = 3/2$ $K_S \to \pi^+ \pi^- $ amplitude.

\item{Fig.3}  CPM graph for $K_L \to \gamma\gamma$ decay.

\item{Fig.4}  CPM graph for $K_S \to \gamma\gamma$ decay.

\item{Fig.5}  L$\sigma$M quark loops (a) and $\pi^+$ loop (b) for $\sigma \to \gamma\gamma$ decay.

\item{Fig.6}  CPM graph for $K_L \to \pi^\circ \gamma\gamma$ decay.

\item{Fig.7}  W-mediated $s \to d$ loop (a) becoming $\Delta I = 1/2$ SQL transition (b)

\item{Fig.8}  Quark $s \to d$ loop representing $K^\circ \to$ vacuum matrix element of weak axial current. 
\end{description}

\setlength{\unitlength}{1cm}

\newpage

\begin{figure}
\begin{picture}(14,20)
\put (0.5,0.0){\epsfxsize=13cm \epsfbox{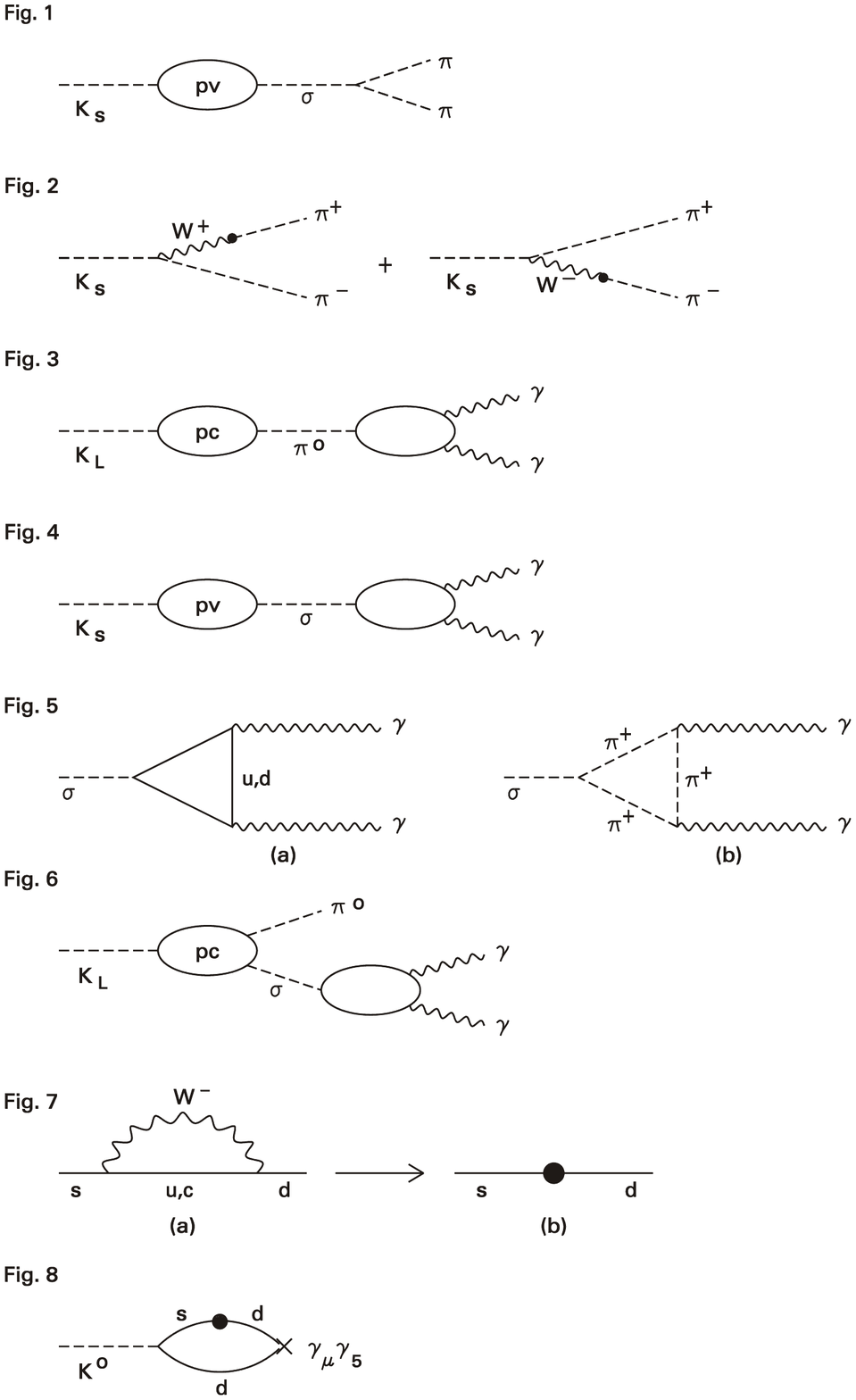}}
\end{picture}
\end{figure}
 

\begin{thebibliography}{99}
\bibitem{18}  Feynman rules normally suggest that fermion and boson loops have opposite sign.
\end{thebibliography}
\end{document}